# Diagram of measurement series elements deviation from local linear approximations

*Lande D.V. (dwl@visti.net), Snarskii A.A. (asnarskii@gmail.com),*
*National Technical University of Ukraine "Kyiv Polytechnic Institute"*

For the detection and visualization of trends, periodicities, local peculiarities in measurement series, the methods of fractal and wavelet-analysis are of considerable current use. One of such methods - DFA (Detrended fluctuation analysis), [1, 2], is used to reveal statistical self-similarity of signals. The essence of this method is as follows. Let there be a measurement series $x_t$, $t \in 1,...,N$. Denote the average value of this measurement series $\langle x \rangle = \frac{1}{N} \sum_{k=1}^{N} x_k$. From the source series, the accumulation series is built:

$$X_t = \sum_{k=1}^{t} (x_k - \langle x \rangle).$$

Then $X_t$ series is divided into time windows of length $L$, linear approximation ($L_{j,L}$) is built according to $X_{k,j,L}$ values from $X_{j,L}$ inside each window (in its turn, $X_{j,L}$ is subset $X_t$, $j = 1,...,J$, $J = N/L$ - the number of observation windows) and a deviation of accumulation series points from linear approximation is calculated:

$$E(j,L) = \sqrt{\frac{1}{L} \sum_{k=1}^{L} (X_{k,j,L} - L_{k,j,L})^2} = \sqrt{\frac{1}{L} \sum_{k=1}^{L} |\Delta_{k,j,L}|^2},$$

where $L_{k,j,L}$ is the value of local linear approximation at point $t = (j-1)L + k$.

Here $|\Delta_{k,j,L}|$ is absolute deviation of element $X_{k,j,L}$ from local linear approximation.

Then the average value is calculated:

$$F(L) = \frac{1}{J} \sum_{j=1}^{J} E(j,L),$$

following which, in case $F(L) \propto L^{\alpha}$, where $\alpha$ is certain constant, conclusions are made on the availability of statistical self-similarity and behaviour of measurement series under study.

Of interest is behaviour of absolute deviation of accumulation series points from linear approximation $|\Delta_{k,j,L}|$ (let us call it $\Delta L$-method) for the real processes, for example, reflecting the intensity of publications on this subject in the Internet. Most commonly, time series corresponding to thematic information flows possess the properties of statistic self-similarity [3], which is confirmed, in particular, by DFA method. Visualization of parameters $|\Delta_{k,j,L}|$ as a function of $t = (j-1)L + k$ and $L$ as a "relief" diagram is of certain interest for studying local peculiarities of the process corresponding to the source measurement series.

Note that division of the source range of values $t \in 1,...,N$ into $J$ nonoverlapping observation windows results in some "inequality" of points inside these windows, which is not principal in case of summation and subsequent approximate estimation, but essential in the analysis of local values and visualization. Therefore, without giving up the idea of linear approximation, it is proposed to choose for each point $t$ such observation window of length $L$, that this point appears to be in its centre (or with a shift at 1 in case of even $L$). Undoubtedly, with regard to this correction, the calculation speed is decelerated $|\Delta_{k,j,L}|$, which is largely compensated by the simplicity of algorithm.

As a time series under study, on which basis the method opportunities will be considered, we shall use a series of daily number of published works on certain subjects in the Internet during a year (Fig. 1). This series was obtained by means of InfoStream content-monitoring system, regularly scanning over 3000 on-line Russian and Ukrainian mass media [4].

"Relief " diagrams obtained as a result of proposed method (this diagram is exemplified in Fig. 2, where lighter tones correspond to larger values $|\Delta_{k,j,L}|$), resemble scalegrams obtained as a result of continuous wavelet-transformations. The fact that dark stripes in the centre of many light coloured areas testify to "stabilization" of large values of series under consideration on a high level is noteworthy.

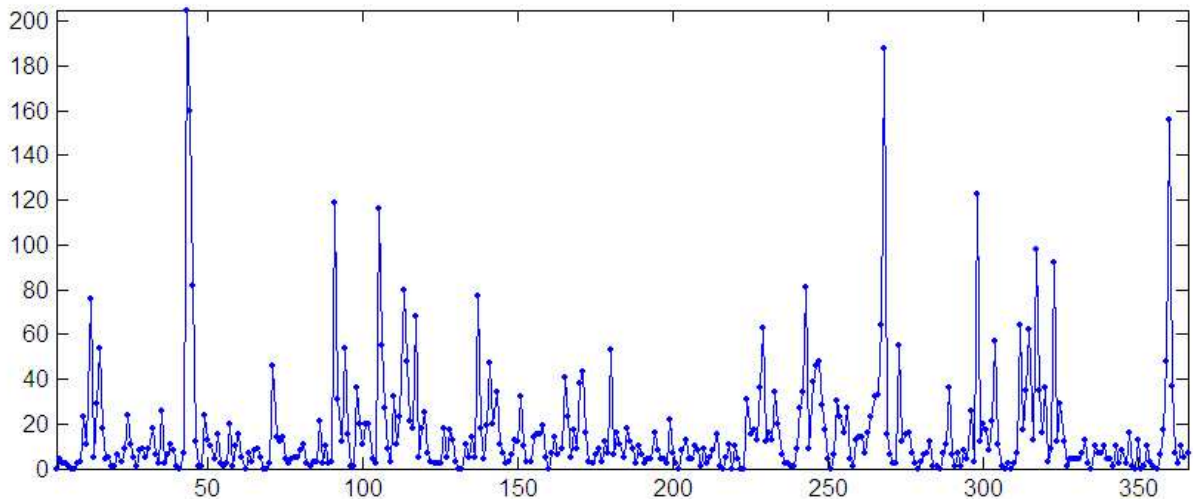

*Fig. 1. Time series of intensity of publications on given subjects*
*(abscissa axis –days of the year, ordinate axis –number of publications)*

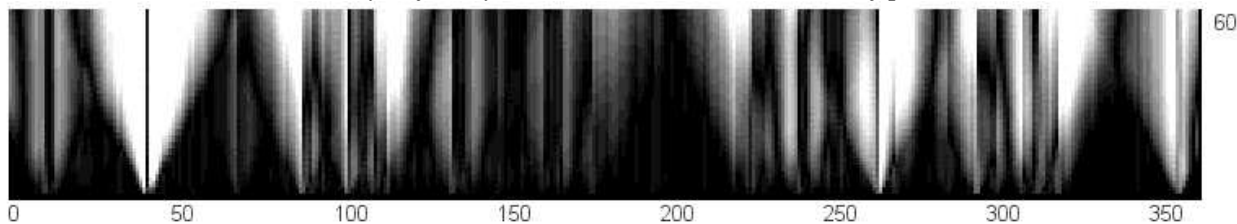

*Fig.2. $\Delta L$-diagram of time series of intensity of subject publications*
*(abscissa axis –days of the year, ordinate axis –size of measurement window)*

Note that $\Delta L$-method proves to be rather efficient for revealing harmonic components of series under study. Fig. 3 shows $\Delta L$-diagram of the series corresponding to sinusoid ($y(i) = \sin(i\pi/7), \ i = 1,...,366$). Application of $\Delta L$-method to the series composed of the number of publications scanned by InfoStream system from Internet without considering subject division, has a pronounced harmonic component (total number of publications depends on the day of the week), that can be seen in Fig. 4. Besides, in this diagram one can see deviations from general dynamics of publication volumes on holidays.

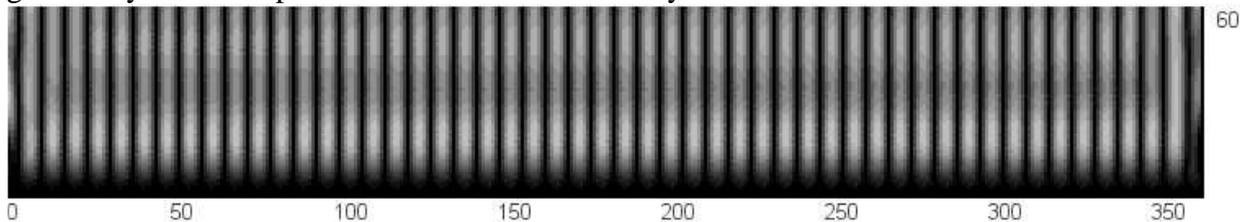

*Fig.3. $\Delta L$-diagram of sinusoid*

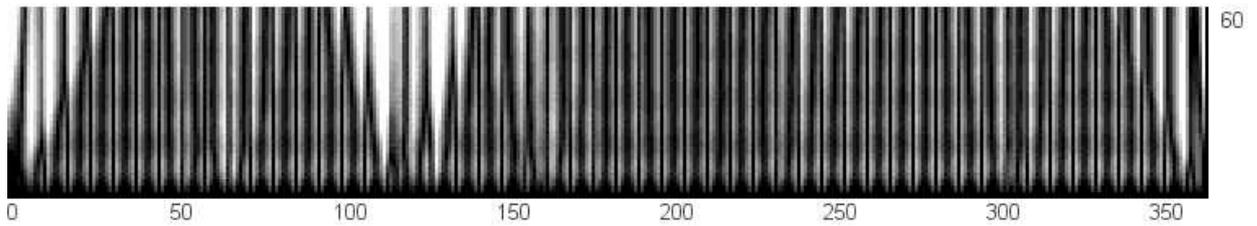

*Fig.4. $\Delta L$-diagram of series of the number of publications scanned daily by InfoStream system in 2008*

$\Delta L$-diagrams are similar in appearance to scalegrams obtained as a result of wavelet-analysis of measurement series. The basic idea of wavelet-transformations lies in the fact that some numerical series, like in the above-mentioned method, is divided into "observation windows", and on each of them there is generated a set of coefficients that are functions of two variables: time and frequency, and thus can be also represented as "relief" diagrams, the so-called scalegrams. In their nature, wavelet-coefficients represent a certain degree of proximity of measurement series under study to certain special function called wavelet [31, 32].

Continuous wavelet-transformation for function $f(t)$ is constructed by means of continuous scale transformations and transfers of wavelet $\psi(t)$ with arbitrary values of scale coefficient $a$ and shift parameter $b$:

$$W(a,b) = (f(t), \psi(t)) = \frac{1}{\sqrt{a}} \int_{-\infty}^{\infty} f(t) \psi^*\left(\frac{t-b}{a}\right) dt.$$

Fig. 5 shows a scalegram – the result of continuous wavelet-analysis (Gauss wavelet) of time series corresponding to process under study.

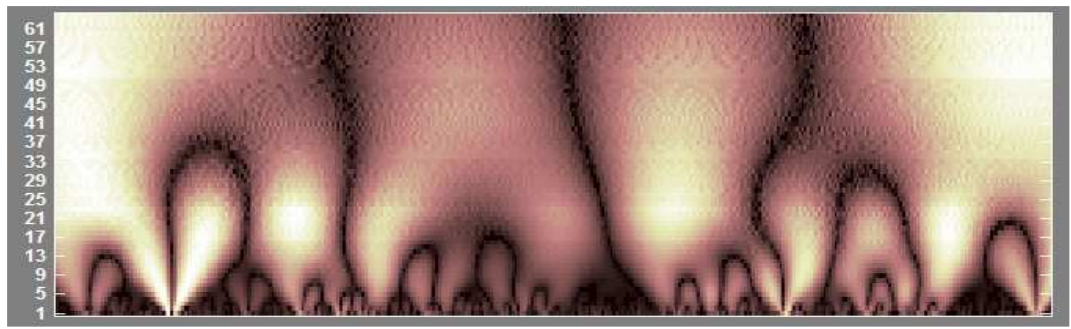

*Fig. 5. Scalegram of time series under study (Gauss wavelet)*

Proposed method for visualization of absolute deviations $\Delta L$, like the wavelet-transformations method, allows (not worse, as it is shown in the example) detecting single and irregular "bursts", drastic changes in the values of quantity figures in different time periods. Note that wavelet-transformations method can be employed with the use of various wavelets. In particular, the use of Haar wavelet (Fig. 6), apparently, is more suitable for analysis of sequence under consideration. However, even the use of Haar wavelet did not allow identifying the peculiarity (local maximum) of the source measurement series during the last days of year 2008, at least, this peculiarity is not shown as skeleton in Fig. 6 б.

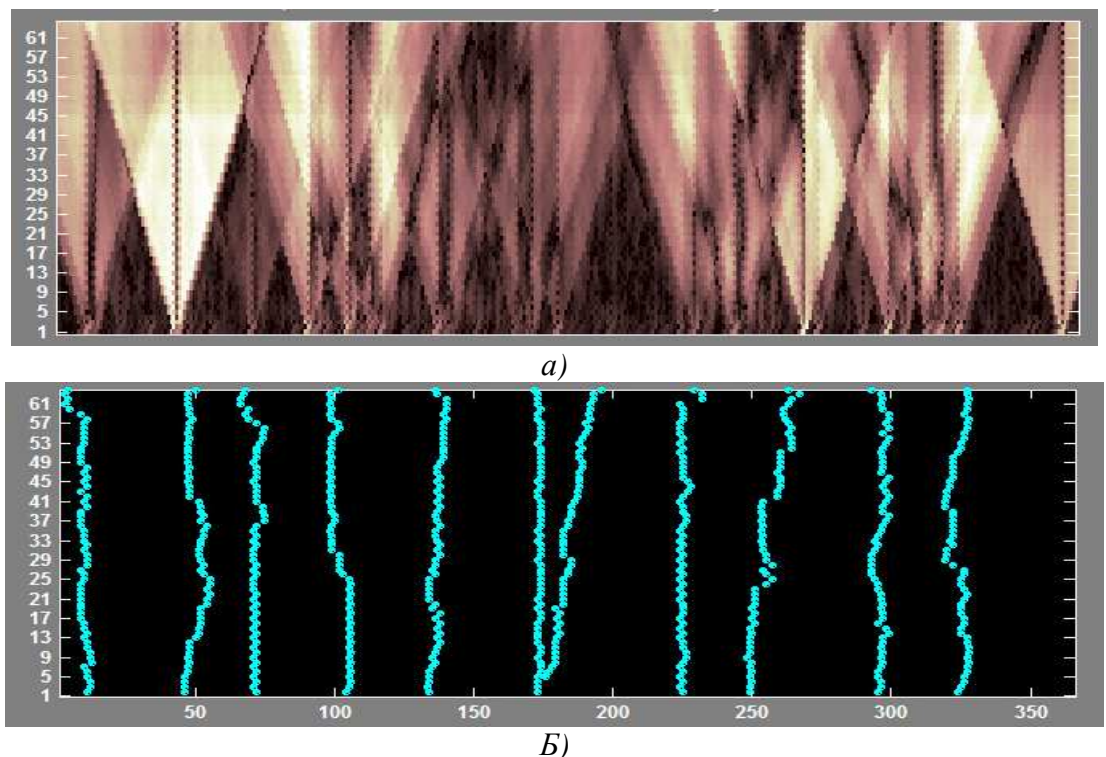

*Fig. 6. Scalegram of time series under study (Haar wavelet):
a) scalegram (abscissa axis – day of the year, ordinate axis - frequency);
б) local maximum lines of scalegram*

$\Delta L$-method that is much easier to realize, nevertheless allowed determining this anomaly. Besides, selection of suitable wavelet for analysis is always a complicated task that need not be solved in case of using $\Delta L$-method. Proposed method is sufficiently easy in program realization and, as experience suggests, can be efficiently used in the analysis of time series in such fields as economics and sociology.

**Abstract**

# Diagram of measurement series elements deviation from local linear approximations

*Lande D.V. (dwl@visti.net), Snarskii A.A. (asnarskii@gmail.com),*
*National Technical University of Ukraine "Kyiv Polytechnic Institute"*

Method for detection and visualization of trends, periodicities, local peculiarities in measurement series ($\Delta L$-method) based on DFA technology (Detrended fluctuation analysis) is proposed. The essence of the method lies in reflecting the values of absolute deviation of measurement accumulation series points from the respective values of linear approximation. It is shown that $\Delta L$-method in some cases allows better determination of local peculiarities than wavelet-analysis. Easy-to-realize approach that is proposed can be used in the analysis of time series in such fields as economics and sociology.

Key words: measurement series, linear approximation, wavelet-analysis, Detrended fluctuation analysis, $\Delta L$-method